\documentclass[aps,onecolumn,superscriptaddress,amsmath,showpacs,tightenlines]{revtex4-1}
\usepackage{graphicx}
\usepackage{subfigure}
\usepackage{natbib}
\usepackage{epsfig}
\usepackage{amsfonts}
\usepackage{amsthm,amsmath,amssymb}
\usepackage{mathrsfs}

\usepackage{epstopdf}
\usepackage{xcolor}
\usepackage[toc,page,title,titletoc,header]{appendix}

\usepackage[utf8]{inputenc}

\begin{document}

\title{Towards a dissipative quantum classifier}
\author{He Wang}
\affiliation{College of Physics, Jilin University,\\Changchun 130021, China}%
\affiliation{State Key Laboratory of Electroanalytical Chemistry, Changchun Institute of Applied Chemistry,\\Changchun 130021, China.}%
\author{Chuanbo Liu}
\affiliation{State Key Laboratory of Electroanalytical Chemistry, Changchun Institute of Applied Chemistry,\\Changchun 130021, China.}
\author{Jin Wang}
\email{jin.wang.1@stonybrook.edu}
\affiliation{Department of Chemistry and of Physics and Astronomy, Stony Brook University, Stony Brook,\\NY 11794-3400, USA}%

\begin{abstract}

In this paper, we propose a novel quantum classifier utilizing dissipative engineering. Unlike standard quantum circuit models, the classifier consists of a central spin-qubit model. By subjecting the auxiliary qubits to carefully tailored strong dissipations, we establish a one-to-one mapping between classical data and dissipative modes. This mapping enables the encoding of classical data within a decoherence-free subspace, where the central qubit undergoes evolution. The dynamics of the central qubit are governed by an effective Lindblad master equation, resulting in relaxation towards a steady state. We first demonstrate the capability of our model to prepare arbitrary single-qubit states by training the inter-coupling of the system and the external dissipations. By elucidating the underlying classification rule, we subsequently derive a quantum classifier. Leveraging a training set with labeled data, we train the dissipative central spin-qubit system to perform specific classification tasks akin to classical neural networks. Our study illuminates the untapped potential of the dissipative system for efficient and effective classification tasks in the realm of quantum machine learning.

\end{abstract}

\maketitle
\section{Introduction}
Machine learning has achieved remarkable success across a broad range of applications in recent years, revolutionizing techniques in both research and industry from image recognition to autonomous vehicles \cite{TMM97,SDS19}. Concurrently, quantum computing has witnessed significant advancements, culminating in experimental demonstrations of quantum supremacy as the latest milestone \cite{MN10,GE19,AF19,HSZ20}. Noteworthy algorithms, including Shor's algorithm \cite{PWS94}, Grover search algorithm \cite{LKG96}, Harrow-Hassidim-Lloyd (HHL) algorithm \cite{AWH09}, quantum approximate optimization algorithm (QAOA) \cite{FE14}, have been proposed, showcasing the potential for quantum advantages over classical counterparts. These breakthroughs have sparked considerable interest in exploring enhanced machine learning models utilizing quantum devices, giving rise to the field of quantum machine learning. The primary objective of quantum machine learning is to expedite machine learning computations by harnessing the speedups afforded by inherent quantum properties, such as entanglement and superposition \cite{ML23}. Notable theoretical advancements and implementations have emerged, including, quantum principal component analyses \cite{LS14}, quantum support vector machines \cite{PR14}, variational quantum eigensolvers \cite{JRM16}, quantum Boltzmann machines \cite{MHA18}, quantum reinforcement learning \cite{DD08,GDP14,ML23b}, quantum feature spaces and kernels \cite{HV19}, and quantum generative adversarial networks \cite{LH19}, among others.

However, it is important to acknowledge that all quantum systems inevitably interact with the environment, leading to dissipation and decoherence, which compromises their quantum nature  \cite{BHP06}.  Mitigating the effects of dissipation and decoherence poses a fundamental challenge in the era of noisy intermediate-scale quantum (NISQ) devices \cite{JP18}. Researchers have devised various techniques to combat decoherence and dissipation, including quantum error correction \cite{JR2019}, decoherence-free subspaces \cite{PZ1997,DAL1998,DAL14}, noiseless subsystems \cite{LV00,MDC06,DAL14}, feedback control \cite{HMW94,ACD00,JZ17}, dynamical decoupling \cite{LV99,DAL14}, the quantum Zeno effect \cite{WMI1990,PF02,KK05,AZC14,AZC16}, and Floquet engineering \cite{CC15,WLY19,SYB20,HW23a}, among others. Interestingly, the environment is not always detrimental to quantum systems; in certain cases, it can even be advantageous. For instance, two initially uncorrelated quantum detectors can become quantum-correlated when both couple to a common field \cite{FB05,BEG13,HW21}, and at times, high temperatures can even amplify quantum correlations \cite{BEG13,HW21}. Even in the realm of quantum machine learning, the environment can play a beneficial role. Dissipation can serve as a valuable resource for universal computation \cite{BK08,FV09,DT19,UK22,UK23,UK23b,GE21,GE22,GE23,GE23b}. Recent work on quantum reinforcement learning \cite{ML23b} shows performance is not significantly impaired by thermal dissipation at sufficiently low temperatures, and can even be enhanced in some cases \cite{ML23}. Now, the concept of dissipative engineering, which harnesses dissipation, is emerging as a powerful tool. It has enabled the preparation of many-body states \cite{FV09,BK08}, and the realization of the exotic phase of matter \cite{SD11,HW23}.

In this paper, we further explore the possibility of realizing dissipative quantum classification tasks by combining dissipative engineering and machine learning architectures. Specifically, we investigate a central spin model (see Fig.\ref{fig1}), commonly employed to describe the hyperfine interaction between an electron spin in a quantum dot and the surrounding nuclear spins  \cite{JS02,MB07}. The auxiliary qubits are subjected to strong tailored dissipation, which can be experimentally realized \cite{NS08,FS14}. Following a short relaxation time, the entire system evolves within a decoherence-free subspace \cite{EMK12,VP18,VP21}. The central qubit, or system qubit, is governed by an effective Lindblad master equation \cite{EMK12,VP18,VP21}. By identifying classical data as the dissipative modes acting on the auxiliary qubits, we can derive a steady state of the central qubit that encodes the classical data in a specific manner. In essence, our dissipative model harnesses engineered dissipation to drive the system into a desired steady state containing the computational output. Through elucidating the classification rule or some others, we may implement certain tasks based on this model. 
we first demonstrate that our model possesses the capability to prepare any desired single qubit state. The state preparation is crucial for quantum computation and quantum communication. Previous studies \cite{BK08,FV09} have proposed the use of dissipative processes to prepare quantum systems in desired states. In this work, we leverage dissipative engineering and tune the coupling within our system to iteratively minimize the loss function, allowing us to prepare any desired single qubit state. Subsequently, we investigate the application of our proposed model to binary classification tasks. Classification is a pivotal branch in machine learning, finding widespread use in both commercial and academic applications such as face recognition, recommendation systems, earthquake detection, and disease diagnosis \cite{TMM97}. Prior works \cite{DT19,UK22,UK23,UK23b} have explored dissipative quantum classifiers based on collision models, where classical data is mapped to the pure state of a collisional qubit. These models require a large ensemble of collisional qubits, serving as information reservoirs, to drive the system qubit to a steady state that encodes the classification result. In our approach, we directly encode the classical data into the steady state of the central qubit system through dissipative drives. This direct relationship between the classical data and the dissipation modes or Lindblad operators can be realized through dissipative engineering \cite{BJ11,MM11,AC17}. Consequently, there is no need to prepare a large ensemble of states. By employing classical machine learning algorithms, we update the coupling between the auxiliary qubits and the central qubit to minimize the cost function. The resulting trained model demonstrates remarkable proficiency in effectively tackling diverse binary classification tasks. 

The remainder of this paper is structured as follows.  In Sec.\ref{model}, we present the model under investigation and describe its evolution. In Sec.\ref{app}, we introduce two applications utilizing the model: state preparation and binary classification. Finally, provide a concluding remarks in Sec.\ref{summary}.

\section{Model and evolution}\label{model}

\begin{figure}[!ht]
    \centering
\includegraphics[width=2.5in]{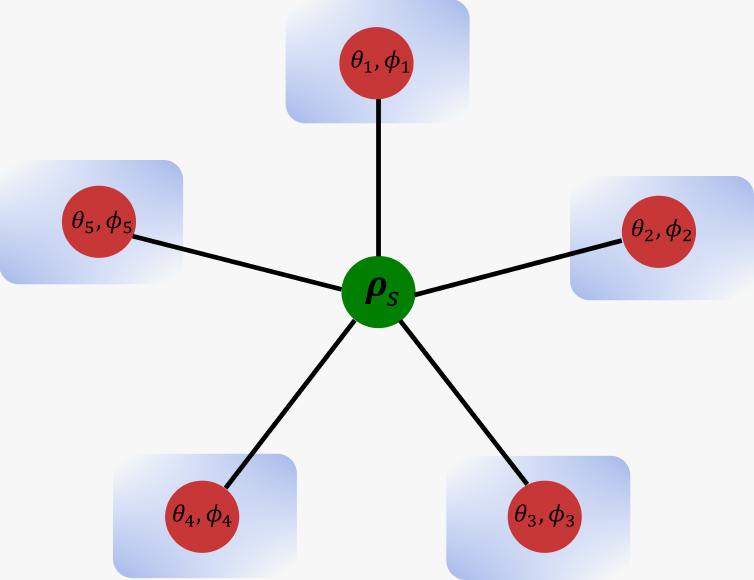}
\caption{\label{fig1} A sketch of the dissipative classifier with five auxiliary qubits. The dissipative classifier incorporates the interaction between the central qubit and the auxiliary qubits, which are subjected to strong engineered dissipative processes. The final computational outcome is encoded within the steady state of the central qubit.}
\end{figure}

In this section, we introduce the model of interest and derive its evolution dynamics. The total system consists of spin qubits, with a central spin qubit coupled to several auxiliary spin qubits. The auxiliary qubits are subjected to engineered strong dissipative processes describable by a Lindblad master equation,

\begin{equation}
    \frac{\partial \pmb \rho(\tau) }{\partial \tau} = -i\left[\pmb H,\pmb \rho(\tau)\right] + \Gamma { \cal \pmb D} [\pmb \rho(\tau)] =-i\left[\pmb H,\pmb \rho(\tau)\right] + \Gamma \sum_{\alpha}\sum_{n=1}^N{ \cal\pmb D}_{\pmb L_n^\alpha} [\pmb \rho(\tau)],
\end{equation}

where $\pmb H$ is the Hamiltonian of the total system under investigation. Additionally, the dissipator $\mathcal{\pmb D}_{\pmb L_n^\alpha}$ is introduced to account for the $\alpha$-th strong dissipative modes affecting the n-th auxiliary qubit. This dissipator follows the standard Lindblad form, defined as $\mathcal{\pmb D}_{\pmb L_n^\alpha} \pmb X =\pmb L_n^\alpha \pmb X \pmb L^{\alpha\dagger}_n - \frac{1}{2} (\pmb L^{\alpha\dagger}_n \pmb L_n^\alpha \pmb X + \pmb X \pmb L^{\alpha\dagger}_n\pmb L_n^\alpha)$. A schematic representation of the system is depicted in Fig.\ref{fig1}.

We select the Lindblad operators as $\pmb L_n^1=\sqrt{\frac{1+\mu}{2}}|s^\perp_n\rangle\langle s_n|$, and $\pmb L_n^2 =\sqrt{\frac{1-\mu}{2}} |s_n\rangle\langle s^\perp_n|$. Here, $|s_n\rangle$ represents an arbitrary vector in the Hilbert space of the n-th qubit, and $\langle s_n|s^\perp_n\rangle=0$. We assume that the dissipator is diagonalizable, and its steady state is non-degenerate. The dissipator's eigenbasis is denoted as $\{\pmb\psi_k^n\}$, corresponding to eigenvalues ${\xi_k^n}$, with $\xi_0^n = 0$. Additionally, we introduce a complementary basis $\{\pmb\varphi_k^n\}$, satisfying the condition $Tr(\pmb\psi_l^n\pmb\varphi_k^m)=\delta_{mn}\delta_{lk}$. Consequently, the eigenbasis of the total dissipator can be expressed as $\pmb\Psi_{\{a,b...\alpha\}}=\pmb\psi_a^1\otimes\pmb\psi_b^2\dots\otimes\pmb\psi_\alpha^N$ with eigenvalue  $\Xi_{\{a,b...\alpha\}}=\xi_a^1+\xi_b^2+\dots\xi_k^n$. Furthermore, we define the complementary basis as $\pmb\Phi_{\{a,b...\alpha\}}=\pmb\varphi_a^1\otimes\pmb\varphi_b^2\dots\otimes\pmb\varphi_\alpha^N$.

In the regime of strong dissipations, characterized by $\Gamma\rightarrow\infty$, the dynamics of the entire system is restricted within a decoherence-free subspace defined by $\pmb \psi_0^1 \otimes \pmb \psi_0^2\dots\otimes\pmb \psi_0^N\otimes \pmb \rho_S(\tau)$ after a short relaxation time. Upon tracing out the auxiliary qubits' degrees of freedom, the evolution of the central spin qubit is described by an effective master equation, as established in previous works \cite{EMK12,VP18,VP21},

\begin{equation}
    \frac {\partial \pmb \rho_S}{\partial \tau} = -i \left[\pmb h_D+ \tilde{\pmb H}_a/\Gamma,\pmb \rho_S \right]+\frac{1}{\Gamma}\tilde{\mathcal{\pmb D}}[\pmb \rho_S]
\end{equation}

where $\pmb h_D=Tr_{\bar{S}} \left( \left(\pmb\Psi_{\{0,0...0\}}\right)\pmb H \right)= Tr_{\bar{S}} \left( \left(\pmb \psi_0^1 \otimes \pmb \psi_0^2\dots\otimes\pmb \psi_0^N\right)\pmb H \right)=Tr_{\bar{0}} \left( \pmb\Psi_{\pmb 0}\pmb H \right)$. The super-operator $Tr_{\bar{S}}(\cdot)$ denotes the operation of tracing out all degrees of freedom except those pertaining to the system. The super-index $\pmb 0$ represents the set $\{0,0,\ldots,0\}$, indicating a specific configuration of eigenvalues.

Leveraging the eigenbasis, we can decompose the total Hamiltonian as

\begin{equation}\begin{split}
    \pmb H &= \sum_{\{a,b,...\alpha\}} \pmb\Phi_{\{a,b...\alpha\}}\pmb g_{\{a,b,...\alpha\}}=\sum_{\{a,b,...\alpha\}} \pmb\Phi_{\{a,b...\alpha\}}^\dagger \otimes \pmb g_{\{a,b,...\alpha\}}^\dagger,\\  
\pmb g_{\{a,b,...\alpha\}} &= Tr_{\bar{0}} ((\pmb I\otimes\pmb\Psi_{\{a,b...\alpha\}}  )\pmb H).
\end{split}\end{equation}

Define the coefficient matrix 
\begin{equation}\begin{split}
C_{\{a,b...\alpha\},\{a',b'...\alpha'\}}&= Tr \left( \pmb\Phi_{\{a,b...\alpha\}}^\dagger \pmb\Phi_{\{a',b'...\alpha'\}} \pmb\Psi_{\pmb 0} \right),
\\ -C_{\{a,b...\alpha\},\{a',b'...\alpha'\}}/\Xi_{\{a,b...\alpha\}}^*&=Y_{\{a,b...\alpha\},\{a',b'...\alpha'\}}=A_{\{a,b...\alpha\},\{a',b'...\alpha'\}}/2 +i B_{\{a,b...\alpha\},\{a',b'...\alpha'\}},
\end{split}\end{equation}
where $A_{\{a,b...\alpha\},\{a',b'...\alpha'\}}=Y_{\{a,b...\alpha\},\{a',b'...\alpha'\}}+Y_{\{a,b...\alpha\},\{a',b'...\alpha'\}}^*$ is a positive matrix and $B_{\{a,b...\alpha\},\{a',b'...\alpha'\}}=(Y_{\{a,b...\alpha\},\{a',b'...\alpha'\}}-Y_{\{a,b...\alpha\},\{a',b'...\alpha'\}}^*)/(2i)$ is a Hermitian matrix. We can derive

\begin{equation}\begin{split}
\tilde{\pmb H}_a &=  \sum_{Re(\Xi_{\{a,b...\alpha\}})<0,Re(\Xi_{\{a',b'...\alpha'\}})<0}  B_{\{a,b...\alpha\},\{a',b'...\alpha'\}}\pmb g_{\{a,b...\alpha\}}^\dagger \pmb g_{\{a',b'...\alpha'\}},\\
 \tilde{\mathcal{\pmb D}} \pmb R &= \sum_{Re(\Xi_{\{a,b...\alpha\}})<0,Re(\Xi_{\{a',b'...\alpha'\}})<0} A_{\{a,b...\alpha\},\{a',b'...\alpha'\}} \left( \pmb g_{\{a',b'...\alpha'\}} \pmb R \pmb g_{\{a,b...\alpha\}}^\dagger - \frac{1}{2} \{\pmb g_{\{a,b...\alpha\}}^\dagger \pmb g_{\{a',b'...\alpha'\}}, \pmb R\} \right)
\end{split}\end{equation}

It is important to note that the aforementioned equation is valid solely under the assumption of strong dissipations. In this regime, as the dissipations acting on the auxiliary qubits increase, the impact of dissipations on the central qubit becomes less pronounced. Nevertheless, it is crucial to acknowledge that dissipations affecting the central qubit cannot be disregarded, as they play a significant role in determining the ultimate steady state of the system \cite{EMK12,VP18,VP21}. 

Now we specify the total Hamiltonian we studied, it reads,

\begin{equation}\label{hamiltonian}
\pmb H =\sum_{n=1}^{N} \vec{\pmb\sigma^0} \cdot (\pmb J_n \vec{\pmb\sigma^{n}}),
\end{equation}

where the $\pmb J_n$ is the coupling matrix between the n-th auxiliary qubit and the central qubit. Owing to the locality of the Hamiltonian, the terms $\pmb g_{\{a,b,...\alpha...\beta...\}} = Tr_{\bar{0}} ((\pmb I\otimes\pmb\Psi_{\{a,b...\alpha...\beta...\}}  )\pmb H)=0$ vanish whenever any two indices $\alpha>0$ and $\beta>0$. Consequently, only the terms $\pmb g_{\pmb k_n} = Tr_{\bar{0}} ((\pmb I\otimes\pmb\Psi_{\{0,0...k...0\}}  )\pmb H)$ may not be zero. Here, the super-index $\pmb k_n$, denotes that only the n-th element is $k$ while the remaining elements in the set of N elements are zero. 

The eigenbasis and corresponding eigenvalues of the superoperator $\mathcal{\pmb D}_{\pmb L_n}$ are as follows,

\begin{equation}\begin{split}
\pmb\psi_0^n &= \frac{1+ \mu}{2}|s_n\rangle\langle s_n| + \frac{1- \mu}{2}|s_n^\perp\rangle\langle s_n^\perp|, \qquad  \xi_0^n=0,   \\
 \pmb\psi_1^n &= |s_n\rangle \langle s_n^\perp|, \qquad  \xi_1^n=-\frac{1}{2},\\
 \pmb\psi_2^n &= |s_n^\perp\rangle \langle s_n|, \qquad \xi_2^n=-\frac{1}{2},\\
 \pmb\psi_3^n &= |s_n\rangle \langle s_n|-|s_n^\perp\rangle \langle s_n^\perp|, \qquad \xi_3^n=-1,
\end{split}\end{equation}

and the complimentary basis are

\begin{equation}\begin{split}
\pmb\varphi_0^n&= \pmb I,\\
  \pmb\varphi_1^n&= |s_n^\perp\rangle \langle s_n|,\\
  \pmb\varphi_2^n&= |s_n\rangle \langle s_n^\perp|,\\
  \pmb\varphi_3^n&= \frac{1- \mu}{2} |s_n\rangle \langle s_n| - \frac{1+\mu}{2}|s_n^\perp\rangle \langle s_n^\perp|.
\end{split}\end{equation}  
 
We can now explicitly compute the coefficient matrix $C_{\{a,b...\alpha\},\{a',b'...\alpha'\}}= Tr \left( \pmb\Phi_{\{a,b...\alpha\}}^\dagger \pmb\Phi_{\{a',b'...\alpha'\}} \pmb\Psi_{\pmb 0} \right)$. The resulting coefficients are as follows: $C_{\pmb 0_m,\pmb 0_n}=\delta_{mn}$, $C_{\pmb 1_m,\pmb 1_n}=\delta_{mn}(1+\mu)/2$, $C_{\pmb 2_m,\pmb 2_n}=\delta_{mn}(1-\mu)/2$, $C_{\pmb 3_m,\pmb 3_n}=\delta_{mn}(1-\mu^2)/4$. All other coefficients are zero. From this, we can easily derive the matrices $A$ and $B$. Finally, the effective master equation for the central qubit is obtained,

\begin{equation}\begin{split}
\tilde{\pmb H}_a &=  0,\\
 \tilde{\mathcal{\pmb D}} \pmb R &= 2(1+\mu)\sum_{n} \left( \pmb g_{\pmb 1_n} \pmb R \pmb g_{\pmb 1_n}^\dagger - \frac{1}{2} \pmb g_{\pmb 1_n}^\dagger \pmb g_{\pmb 1_n} \pmb R-\frac{1}{2} \pmb R \pmb g_{\pmb 1_n}^\dagger \pmb g_{\pmb 1_n} \right)+2(1-\mu)\sum_{n} \left( \pmb g_{\pmb 1_n}^\dagger \pmb R \pmb g_{\pmb 1_n} - \frac{1}{2} \pmb g_{\pmb 1_n} \pmb g_{\pmb 1_n}^\dagger \pmb R-\frac{1}{2} \pmb R \pmb g_{\pmb 1_n} \pmb g_{\pmb 1_n}^\dagger \right)+\\&\quad+(1-\mu^2)/2\sum_{n} \left( \pmb g_{\pmb 3_n} \pmb R \pmb g_{\pmb 3_n}^\dagger - \frac{1}{2} \pmb g_{\pmb 3_n}^\dagger \pmb g_{\pmb 3_n} \pmb R-\frac{1}{2} \pmb R \pmb g_{\pmb 3_n}^\dagger \pmb g_{\pmb 3_n} \right),\\
\pmb h_{D} &= \sum_{n=1}^{N} \pmb g_{\pmb 0_n}.
\end{split}\end{equation}

Here, we analyzed a special open central spin model. In the following, we will showcase how this system can achieve the classification by tailoring dissipations. The classification result is encoded in the steady state of the central qubit.  

\section{Applications}\label{app}

In the context of the previously presented model, we aim to employ it for quantum machine learning tasks. Initially, we showcase its capability to prepare arbitrary single-qubit states. Subsequently, we utilize the model for classification purposes. An essential aspect of quantum machine learning algorithms is the encoding of classical data into a quantum system or state. This step is crucial as it directly impacts the feasibility of the overall task. Conventionally, classical information is encoded into a quantum circuit, and the ground state is allowed to evolve through the circuit. However, we propose a distinct approach in which we encode the classical data into the steady state of the central qubit through environmental engineering. To achieve this, we parameterize the states $|s_n\rangle$ and $|s_n\rangle$ as follows,

\begin{equation}\label{evl}
|s_n\rangle=\left(
\begin{array}{c}
      \cos(\theta_n/2) e^{-i \phi_n/2}
      \\
      \sin(\theta_n/2) e^{i \phi_n/2}
\end{array}\right),\qquad |s_n^\perp\rangle=\left(
    \begin{array}{c}
      \sin(\theta_n/2) e^{-i \phi_n/2}
      \\
      -\cos(\theta_n/2) e^{i \phi_n/2}
    \end{array}
  \right).
\end{equation}
By tailoring dissipations on the n-th auxiliary qubit, we can encode the classical data $\theta_n$ and $\phi_n$ into the system, as illustrated in Fig. \ref{fig1}. Notably, while classical machine learning employs N classical bits as N inputs, our model accommodates $2^N$ inputs with N qubits. Consequently, our model exhibits an exponentially larger computational space compared to its classical counterpart. The strong dissipations at play ensure that the system evolves within the decoherence-free subspace  $\pmb \psi_0^1(\theta_1,\phi_1) \otimes \pmb \psi_0^2(\theta_2,\phi_2)\dots\otimes\pmb \psi_0^N(\theta_N,\phi_N)\otimes \pmb \rho_S(\tau)$ after a transient relaxation process.
As a result of this encoding process, the central qubit eventually reaches a steady state denoted as $\pmb \rho_S(\theta_1,\phi_1,\theta_2,\phi_2,...\theta_N,\phi_N)$. Hence, through the careful design of dissipations, we can effectively encode classical data into the steady state of the central qubit. Notably, unlike the dissipative classifier based on the collision model discussed in previous works such as \cite{DT19,UK22,UK23,UK23b}, our approach eliminates the need to prepare a large ensemble of auxiliary qubits in the same state to store classical data. Additionally, our scheme offers the advantage of not requiring an initialization of both the auxiliary qubits and the central qubit in specific states. The system naturally relaxes to a unique steady state after each measurement. Consequently, our approach exhibits improved experimental feasibility compared to alternative methods.

By introducing the orthogonal vectors  $\vec{v_n}(\theta_n,\phi_n) = (\sin \theta_n \cos \phi_n, \sin \theta_n \sin \phi_n, \cos\theta_n )$, $\vec{v_n'} = \vec{v_n} \left(
  \frac{\pi}{2}-\theta_n,\phi_n+\pi \right)$, $\vec{v_n''} = \vec{v_n} \left(\frac{\pi}{2},\phi_n+\frac{\pi}{2} \right)$, we obtain the following expression, 
  
\begin{equation}\begin{split}
\pmb g_{\pmb 0_n}&=\mu(\pmb J_n\vec{v_n}) \cdot \vec{\pmb\sigma^0}\\
\pmb g_{\pmb 1_n}&=(\pmb J_n\vec{v_n'}) \cdot \vec{\pmb \sigma^0} - i (\pmb J^n\vec{v_n''}) \cdot\vec{\pmb \sigma^0},\\
\pmb g_{\pmb 2_n}&=\pmb g_{\pmb 1_n}^\dagger\\
 \pmb g_{\pmb 3_n}&= 2 (\pmb J_n\vec{v_n}) \cdot \vec{\pmb\sigma^0}.
\end{split}\end{equation}

One point that needs clarification is the rationale behind the need for strong dissipation. While it is true that weaker dissipation could, in principle, achieve the same objective if the steady state of the central qubit is unique, it is important to consider the possibility of multiple steady states or non-steady states emerging \cite{VVA14}. The final states may depend on the initial states, which significantly impedes the extraction of information from the final states and adversely affects the regular operation of the classifier. In addition, it is essential to consider the evolution of the entire system rather than simply focusing on the evolution equation (Eq.\ref{evl}) of the central qubit. This can pose challenges in analyzing systems with a large number of auxiliary qubits. %Lastly, if the engineering dissipation is relatively small to the extent that the background noise in the real environment cannot be neglected, it will further disturb the operation of the system. 
Therefore, it becomes crucial to employ strong dissipation to ensure a unique steady state and facilitate the analysis and utilization of the system. Moreover, it is worth noting that the specific forms of the Lindblad operators in our scheme are not unique and can be chosen based on the experimental constraints and feasibility. The key criterion is that the dissipator $\mathcal{\pmb D}_{\pmb L_n^\alpha}$ must be guaranteed to be diagonalizable and non-degenerate in the steady state. As long as this condition is fulfilled, our scheme remains effective for achieving the desired outcomes.

\subsection{Quantum state preparation}

In this subsection, we demonstrate that our model is capable of preparing arbitrary single-qubit states. For the sake of simplicity, we will concentrate on two auxiliary qubits for the remainder of this paper. Despite the relatively small number of auxiliary qubits in the system, it still offers considerable functionality. To prepare the desired target state, we employ an iterative process where we update the system parameters and dissipation modes. This iterative update aims to minimize the difference between the target state and the current steady state, progressively reducing the discrepancy. To quantify this difference, we introduce a loss function

\begin{equation}\label{loss1}
L(\pmb J_1,\pmb J_2, \theta_1,\phi_1, \theta_2,\phi_2)=\sqrt{Tr(\pmb \sigma_x (\pmb \rho_{S}-\pmb \rho_{desired}))^2+Tr(\pmb \sigma_y (\pmb \rho_{S}-\pmb \rho_{desired}))^2+Tr(\pmb \sigma_z (\pmb \rho_{S}-\pmb \rho_{desired}))^2}.
\end{equation}

Upon closer examination, we observe that Eq.\ref{loss1} evaluates to $0$ if and only if $\pmb \rho_{S}=\pmb \rho_{desired}$. To update the parameters, we employ the gradient descent algorithm. This involves updating both the coupling matrix $\pmb J_n$ and the dissipative parameters $\theta_n, \phi_n$ according to the following equations: $\pmb J_n=\pmb J_n-\eta \frac{\partial L}{\partial \pmb J_n}$  and  $\theta_n= \theta_n-\eta \frac{\partial L}{\partial \theta_n}$, $ \phi_n= \phi_n-\eta \frac{\partial L}{\partial \phi_n}$. Here, $\eta$ denotes the learning rate used in the update process.

\begin{figure}[!ht]
    \centering
\includegraphics[width=5in]{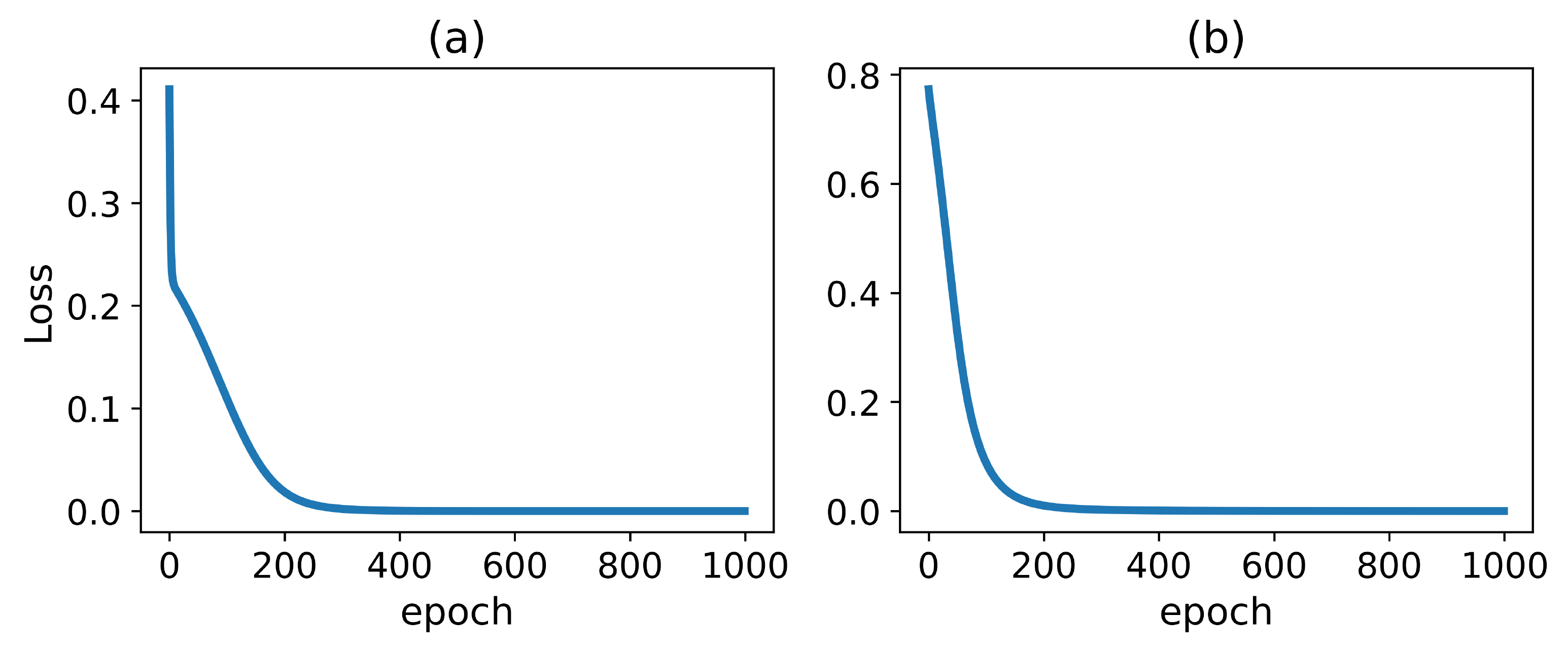}
\caption{\label{fig2} The trend of the loss function as the number of epochs increases The preparation of (a) single-qubit random states, and (b) the coherent state $\frac{1}{\sqrt{2}}(|0\rangle+|1\rangle)$. Throughout the process, we employ a constant learning rate of $\eta=0.05$. The remaining parameters are set as $\mu=1$ and $\Gamma=100$.}
\end{figure}

As previously stated, our model, which incorporates both system and auxiliary qubits, does not require initialization to a specific state. This characteristic of the model presents significant advantages in practical applications. The process of state preparation involves waiting for the system to converge to the steady state, measuring the current loss function, and subsequently updating the parameters based on the loss function gradient. In Fig.\ref{fig2}, we illustrate the variations of the loss function for two different scenarios. In Fig.\ref{fig2}(a), the desired state is a randomly selected single-qubit state, while in Fig.\ref{fig2}(b), the desired state is the maximal coherent state $\frac{1}{\sqrt{2}}(|0\rangle+|1\rangle)$. Upon closer examination, we observe similar trends in both scenarios. The loss function decays monotonically and rapidly approaches zero as the training epochs increase. This indicates that, after a certain number of training epochs, the desired state is achieved. Consequently, this model can be employed as a preparator for single-qubit states.

\subsection{Dissipative quantum classifier}

In this subsection, we present the capability of our proposed model to carry out binary classification tasks. Specifically, we set $\phi_1=\phi_2=0$ and only focus on the classification of $\theta_1$ and $\theta_2$. As previously mentioned, we establish a one-to-one mapping between classical data and the dissipative modes. This mapping allows for the encoding of classical information onto the steady state of the system qubit. The computational result can be obtained by measuring certain observables. By defining classification rules, we can utilize a large dataset with labeled data to train the Hamiltonian (Eq. \ref{hamiltonian}) of the system. Consequently, the system becomes capable of classifying the specific problem at hand. To facilitate this process, we employ the Logistic regression algorithm. Logistic regression is widely employed for supervised neural networks and classification tasks involving discrete outcomes. It serves as a generalization of linear regression, with adaptations made to suit classification tasks. We define the activation function $f$ as follows:

\begin{equation}
f(\langle\pmb \sigma_z\rangle)=\frac{1}{1+\exp{(-k\langle\pmb \sigma_z\rangle})},
\end{equation}

which corresponds to the Sigmoid function and $k$ is a control parameter. The resulting value of $f(\langle\pmb \sigma_z\rangle)$ represents the probability of classifying the data into class 1. Assuming that if $f(\langle\pmb \sigma_z\rangle)\geq 0.5$ (or equivalently, $\langle\pmb \sigma_z\rangle\geq 0$), the data will be classified as class 1, otherwise, it belongs to class 0. To quantify the difference between the actual labels and the training results, we employ a commonly used cost function:
\begin{equation}
C(\pmb J_1,\pmb J_2)=-\frac{1}{N}\sum_i^N y_i(\theta_1^i,\theta_2^i)\log{\left( f(\langle\pmb \sigma_z\rangle)(\theta_1^i,\theta_2^i)\right)}+(1-y_i(\theta_1^i,\theta_2^i))\log{\left(1-f(\langle\pmb \sigma_z\rangle)(\theta_1^i,\theta_2^i)\right)},
\end{equation}

Here, $N$ denotes the number of samples in the training set, and $y_i$ is the label of the data $(\theta_1^i,\theta_2^i)$. We employ the gradient descent algorithm to update the parameters $\pmb J_n$. The parameter update is given by  $\pmb J_n=\pmb J_n-\eta \frac{\partial C}{\partial \pmb J_n}$, where $\eta$ represents the learning rate. In the two-dimensional data space, we generate two synthetic datasets known as class 0 and class 1, distinguished by the following features:

\begin{align}
\begin{cases}
class\quad 1, &  \theta_2 \geq g(\theta_1) \\
class\quad 0, &   \theta_2 < g(\theta_1)
\end{cases}
\end{align}
where $g(x)$ is the decision boundary. The form of $g(x)$ is directly related to a concrete problem.

\begin{figure}[!ht]
    \centering
\includegraphics[width=7in]{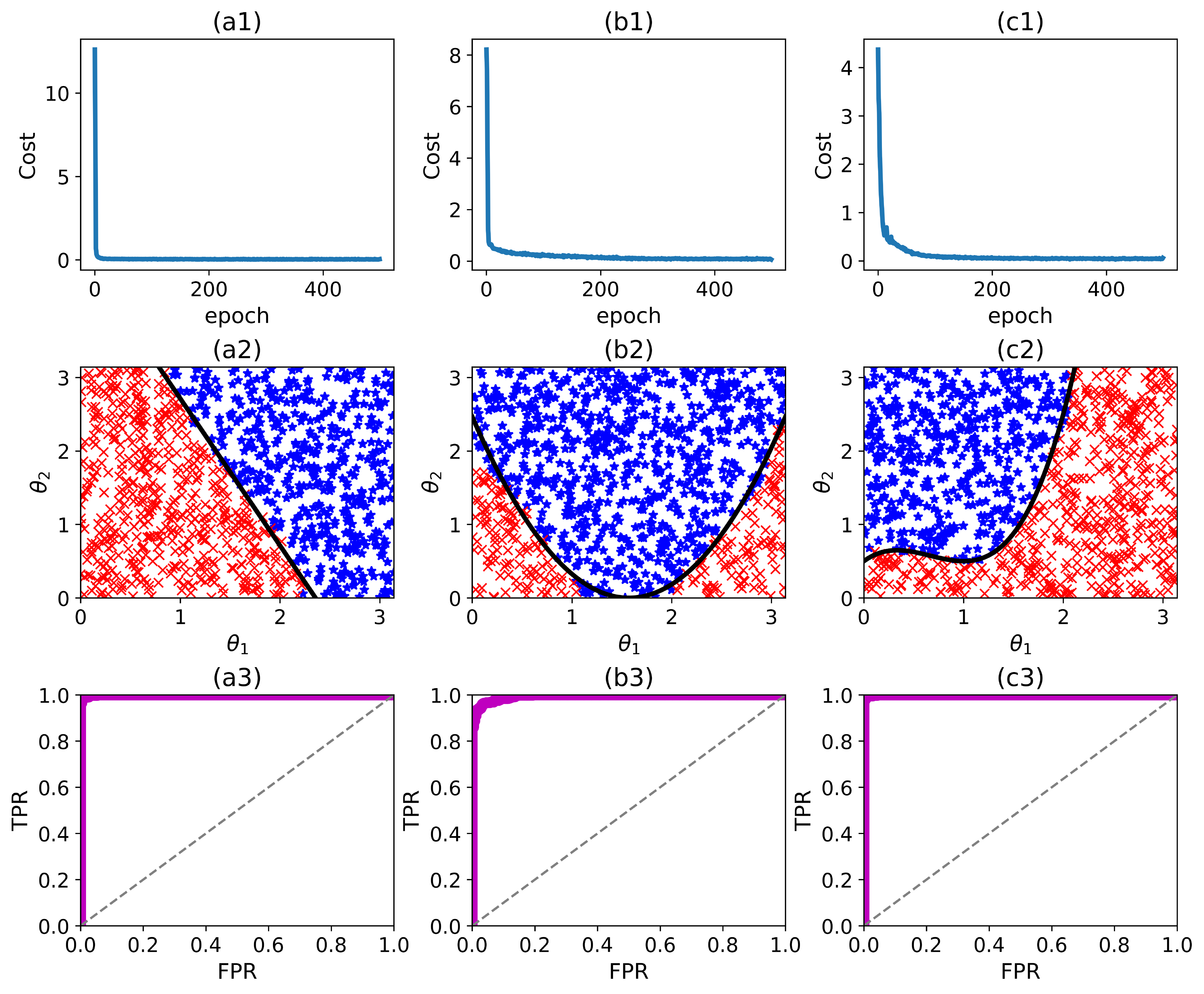}
\caption{\label{fig3} The trend of the cost function with respect to the number of epochs for decision boundary (a1) $g(x)=-2x+1.5\pi$, (b1) $g(x)=x^2-\pi x+\frac{\pi^2}{4}$ and (c1) $g(x)=x^3-2 x^2+x-0.5$. The corresponding validation results are shown in subplots (a2), (b2), and (c2). In these subplots, the blue star represents class 1, while the red cross represents class 0. We utilize a cosine annealing learning rate $\eta$ for training. The ROC curve is also plotted in (a3,b3,c3). The gray dotted line represents the performance of the random classifier that does not possess any discriminatory power. The learning rate starts with a maximum value of $\eta_{\text{max}}=0.05$ and gradually decays to the minimum value of $\eta_{\text{min}}=0.001$ at the final epoch. Other parameters used in the training process include $\mu=1$ and $\Gamma=100$.}
\end{figure}

In Figure \ref{fig3}(a1), we observe the variation of the cost function as the number of epochs increases for the linear decision boundary. The cost function rapidly decreases to a negligible value. Achieving a minimum value for the cost function indicates that the trained model closely approximates the true distribution. This observation is confirmed in Figure \ref{fig3}(a2), where we randomly select points from the data space and classify them using the trained model. The results generated by the model are almost indistinguishable from the actual data, with an accuracy of up to $98.7\%$. Furthermore, we investigate non-linear classification tasks in Figures \ref{fig3}(b1) and \ref{fig3}(c1). The cost functions also exhibit fast decay, reaching a plateau with minor fluctuations. These fluctuations can potentially be mitigated by employing more sophisticated schemes. Although the cost function does not reach its minimum value and exhibits small fluctuations, the current results demonstrate that the trained model is highly effective for classification purposes. We again uniformly and randomly sample points from the data space and classify them using the trained model. In Figures \ref{fig3}(b2) and \ref{fig3}(b3), the label distribution provided by the trained model aligns well with the boundaries corresponding to the real data. The achieved accuracies are $96.2\%$ and $98.9\%$, respectively. In order to further characterize the classification performance of our model, we plot the receiver operating characteristic curve (ROC curve) in Figure \ref{fig3}(a3,b3,c3). The ROC curve illustrates the trade-off between the true positive rate (TPR) and the false positive rate (FPR) as the discrimination threshold of a classifier is varied. By plotting the ROC curve, we gain valuable insights into the classification performance of our model. A poorly performing or random classifier would yield a diagonal line from the bottom-left to the top-right corner, indicating that the true positive rate is similar to the false positive rate across all thresholds. Conversely, an effective classifier would exhibit an ROC curve closer to the top-left corner of the plot, indicating a high sensitivity to correctly identifying positive instances while maintaining a low false positive rate for negative instances. Remarkably, this is precisely the case demonstrated in Figure \ref{fig3}(a3, b3, c3). Additionally, the area under the ROC curve (AUC) is often calculated to summarize the overall performance of the classifier. A higher AUC value indicates better discrimination ability, with a perfect classifier achieving an AUC of 1. In our model, the AUC values for Figure \ref{fig3}(a3, b3, c3) are 0.9995, 0.9950, and 0.9997, respectively. Hence, our proposed model is fully capable of performing binary classification tasks.

\section{Summary}\label{summary}

In this work, we propose a quantum machine learning scheme based on dissipative engineering, utilizing a central star model. By optimizing the coupling between the auxiliary qubit and the central qubit, along with the dissipation modes, we can leverage classical machine learning algorithms to prepare any desired single qubit state. Given classical data, we establish a one-to-one mapping to these engineered dissipative modes. By specifying classification rules in advance, the classification results are encoded in the steady state of the central qubit. The classification can be extracted by measuring specific observables. Our results demonstrate a high accuracy rate, highlighting the exceptional capability of our proposed model in performing diverse binary classification tasks. Expanding the system beyond a single qubit to include multiple qubits has the potential to unlock additional functionalities, as entanglement can emerge. Furthermore, our model may also find applications in preparing arbitrary many-body states \cite{VP20}. Exploring these avenues is a focus of our future work.

\section{Acknowledgement}
H. W. and C.-B.L. acknowledge the support of the National Natural Science Foundation of China, Grant (No. 21721003, No. 12234019 and No.32000888) and the support of the Jilin Province Science and Technology Development Plan Grant 20230101152JC.

\end{document}